\documentclass[fleqn,twoside]{article}
\usepackage{gc}
\usepackage{amssymb}

\heads{Vladimir V. Kassandrov}
      {On Hierarchy and Equivalence of Relativistic Equations for Massive Fields}

\newcommand{\be}{\begin{equation}\label}
\newcommand{\ee}{\end{equation}}
\newcommand{\prt}{\partial}
\newcommand{\p}{\prime}
\newcommand{\bib}{\bibitem}

\begin{document}
\twocolumn[

\Title{ON HIERARCHY AND EQUIVALENCE        \yy
       OF RELATIVISTIC EQUATIONS FOR MASSIVE FIELDS}

%%%          EITHER

   \Author{Vladimir V. Kassandrov\foom 1            %%
              }   %%   If there is
          {Institute of Gravitation and Cosmology, Peoples' Friendship University of Russia, 6 Mikluho-Maklay St., \\ Moscow 117198, Russia}              %%   a single address

%%%          OR

\Abstract
    {A non-canonical correspondence of the complete set of 
solutions to the Dirac and Klein-Gordon free equations in Minkowski 
space-time is established. 
This allows for a novel viewpoint on the relationship of relativistic 
equations for different spins and on the origin of spinor transformations. 
In particular, starting from a solution to the Dirac 
equation, one obtains a chain of other solutions to both Dirac and Klein-Gordon equations. 
A comparison with the massless case is performed, and examples of non-trivial 
singular solutions are presented. A generalization to Riemannian space-time and inclusion of 
interactions are briefly discussed.}

]  %%%%%%%%%%%%%%%%%%%%%%%%%%%%%  End of temporary one-column mode
\email 1 {vkassan@sci.pfu.edu.ru}

\section{Introduction}

It is generally accepted that the linear relativistic field equations are 
mutually independent and in a one-to-one correspondence to 
the finite-dimensional (tensorial or spinorial) irreducible representations of the 
Lorentz (Poincar\'e) group [1,2]. In quantum field theory, 
they describe particles of integer or half-integer spin, respectively. 
he correspondence between two observable types of elementary particles, bosons 
and fermions, and the representations of the invariance group of  
physical space-time is certainly one of the most elegant and 
trustworthy relationships established in physics in general.

Spinor fields, especially the Dirac spin-1/2 field, play a principal role 
at the {\it microlevel} being related to almost all stable elementary particles,  
the proton, electron and neutrino (or to quarks and leptons at a deeper level). 
However, in cosmology the role of spinors is generally considered to be restricted. 
There are a rather limited number of works dealing with cosmological scenarios  
that involve some (nonlinear or interacting) spinor fields invoked to avoid the 
primordial singularity [3], to ensure the inflationary [4] or accelerated [5] 
cosmological regimes etc.

However, the spinor field is in fact unable to represent the vacuum structure 
due to non-invariance of spinors under Lorentz transformations. Instead, 
{\it scalar fields} naturally perform the role of a physical vacuum both at 
the microlevel (Higgs fields) and in the Universe as a whole 
(inflaton fields etc.).

In fact, it looks an enigma why such fields, all originating from  
representations of the same Lorentz group, are so distinct in their manifestations 
in nature? It looks a mystery why there are no {\it stable} particles of zero 
spin described by the Klein-Gordon equation. Or, why there is no massless 
spin-1/2 particle (if one accepts the concept of a {\it massive} neutrino)
corresponding to such a fundamental and fairly simple equation as the Weyl one?

All these and other similar questions, of both pragmatical or ``methaphysical'' 
nature, force one to return back again to the problem of a unified description 
of all elementary particles (irrespective of spins and masses), in particular, 
through some ``hidden'' universal structure of field equations. 
In the past and even in the recent times, there have been  
a number of attempts to bind together the formal structures of distinct 
relativistic equations and the  sets of their solutions or, in other words, 
{\it to prove in a sense their full equivalence} to one another [6,7] (see also [8,9] 
and references therein). 
There are some motivations that stimulate these attempts, the most evident of 
them being the possibility of representing the whole set of {\it wave 
(massless) equations} in a universal 2-spinor form [10-13]. For example, the 
{\it Weyl equation} for a 2-spinor $\{\xi_A\},~~A=1,2$ of a massless spin-1/2 
particle 
\be{weylS}
\prt^{AA^\p} \xi_{A} = 0,
\ee
has a structure quite similar to that of the {\it Maxwell equations} for a 
(symmetric) spinor $\varphi_{(AB)}$ of the electromagnetic field strength:
\be{maxwell}
\prt^{AA^\p} \varphi_{(AB)}=0
\ee
though the latter corresponds to a particle of {\it integer} spin, the photon. 
Though this formal similarity does not imply any equivalence of the two equations 
with the respect of their solutions, in fact this is really the case. For free fields 
in Minkowski space-time, this has been proved, in particular, in our works [14] 
(see also [6]). Moreover, according to the results presented therein, 
{\it any (regular or singular) solution of both Weyl and Maxwell equations may be obtained (via consequtive 
differentiation) from a solution of the ordinary wave equation for a  
one-component complex scalar field} $\phi\in \mathbb {C}$:
\be{scalar}
\Box \phi = 0
\ee
Of course, the converse statement is also true, so that we arrive at some 
informal equivalence between the sets of massless equations for different 
(0, 1/2  and 1) spins in the above formulated sense. 

The result announced is quite nontrivial.  On can note, in particular, that, 
owing to this equivalence, it is possible, say, to write out an analogue of the 
Coulomb solution for the Weyl equation [14], to equip this field distribution 
with at least {\it two energy-like densities}  
(the first one being the canonical indefinite density of the 2-spinor field 
whereas the second one is the positive-definite density inherited from the 
structure of the Maxwell field) etc. Numerous aspects of the equivalence 
obtained and powerful algebraic methods for generating solutions of wave 
equations in the massless case can be found in [14].

As for the more refined case of equations with nonzero mass, it is well known 
that the Klein-Gordon second order equation may be represented in the form of 
the {\it Duffin-Kemmer} first-order matrix equation for a 5-component wave 
function (composed from the initial scalar field along with its 4-gradient). 
However, this construction does not testify to the equivalence of these 
equations and, all the more, has  nothing in common with the Dirac field. 
On the other hand, 
any component of the Dirac bispinor identically satisfies the Klein-Gordon equation.   
It is commonly accepted that, generally, the converse is not true, so that the 
structure of the Dirac equation is more rigid than that of, say, the 
Klein-Gordon equation for a ``4-column'' field (digressing here from their 
transformation properties). However, in this article we show that the situation 
is much more interesting and nontrivial. Namely, we demonstrate (Section 2) that 
the 4-component Klein-Gordon field acts as {\it a field of potentials} for any 
Dirac bispinor. The hidden gauge invariance of the Dirac equation with respect 
to special trasformations of their ``Klein-Gordon potentials'' is also considered.

Thus we demonstrate that {\it any solution to the Dirac equation can be obtained 
(by differentiation) from some solution of a 4-multiplet of the complex 
Klein-Gordon fields and vice versa}. In this sense, these two equations may be 
considered to be equivalent, so that {\it the Dirac 
equation is nothing but a set of four particular constraints on the derivatives 
of a 4-component Klein-Gordon field!} 

A lot of problems of interest naturally arise in this context. One of them is 
a curious possibility of generating a number of (in general, functionally 
independent) solutions to the Dirac and Klein-Gordon equations starting from an 
arbitrary Dirac bispinor. As a result, we arrive at a natural hierarchy of 
solutions to both equations. The chain of solutions resembles a similar chain 
of solutions to the Weyl equation related to solutions of d'Alembert and Maxwell 
equations. This construction is discussed in Section 3 where some remarkable 
examples of singular solutions to massless and massive equations are also presented. 

Section 4 is devoted to the problem of origin of the spinorial transformation law 
for the Dirac field when the latter is obtained from a 4-component 
Klein-Gordon ``scalar'' field. Section 5 concludes the consideration. 

To simplify the exposition, we do not make use of the more convenient 2-spinor 
formalism and operate only with the standard $\gamma$-matrix representation of 
the Dirac equation. Only the simplest case of free fields in the Minkowski flat 
background with the metric $\eta^{\mu\nu}$ 
(of the $+,-,-,-$ signature) is considered, and possible generalizations are 
briefly disscussed only in the final Ssection 5. As usual, we accept the system 
of units such that $c=\hbar=1$.

\section{The Klein-Gordon ``potentials'' for solutions of the Dirac equation}

Consider a multiplet of 4 complex fields $\phi=\{\phi_a\},~~a=1,2,3,4$, each 
subject to the Klein-Gordon equation. In a 4-column representation we have
\be{column}
(\Box - m^2)\phi = 0 ,
\ee
$\Box := -\eta^{\mu\nu}\prt_\mu \prt_\nu$ being the d'Alembert wave operator and $m$ the 
(common) mass of a ``quantum'' of the $\phi$-fields. The Klein-Gordon operator may    
be factorized into a product of two commuting first-order Dirac operators 
$D,D^*$:
\be{factor}
(\Box - m^2) = DD^* = D^*D,
\ee
\be{oper}
D:= i\gamma^\mu \prt_\mu - m,~~~D^*:=i\gamma^\mu \prt_\mu + m,
\ee
where $\gamma^\mu$ are the four canonical $4\times 4$ Dirac matrices    
satisfying the commutation relations
\be{commutate}
\gamma^\mu \gamma^\nu + \gamma^\nu \gamma^\mu = 2\eta^{\mu\nu}.
\ee

Let us now define another 4-component complex field $\chi$ through  
derivatives of the initial $\phi$ as follows:
\be{poten}
\chi:=D^*\phi
\ee
Then, as a consequence of (\ref{column}) and (\ref{factor}), this field 
identically satisfies the Dirac equation
\be{dirac}
D\chi = DD^*\phi \equiv 0.
\ee

Thus any solution to the 4-component Klein-Gordon equation gives rise to a 
solution to the free Dirac equation. Note that the components themselves may be 
functionally independent or not; in particular, some of them may be identically 
equal to zero. Of course, we here deal with solutions differentiable in a connected domain of the Minkowski 
space. However, they are not necessary regular everywhere (being then a composition 
of plane waves); on the contrary, they may be singular at boundary points, in particular, 
contain poles or branching points. The corresponding examples (for the case of 
massless fields) may be found in [14-17] while for massive case they will  
be presented below (Section 3).

Conversely, let now some solution $\chi$ to the Dirac equation (\ref{dirac}) is 
given. It is well known that any component of the Dirac field satisfies the 
Klein-Gordon equation
\be{ident}
0=D^*(D\chi) = (\Box-m^2)\chi \equiv 0.
\ee
However, we are interested in another thing, namely, in restoration of the 
generating {\it potentials} $\phi$ from the basic relation (\ref{poten}). 
This is a system of four inhomogeneous first-order PDE's which, for any given 
$\chi$ in the l.h.s., can always be (locally) resolved with respect to the potentials $\phi$ 
(just as is the case for the field strengths and potentials of the electromagnetic 
field). Of course, the potentials obtained will obey the Klein-Gordon 
equation
\be{ident2}
0=D\chi = DD^*\phi=(\Box-m^2)\phi \equiv 0.
\ee

As expected, the potentials are defined non-uniquely, up to the general solution 
of the ``homogeneous Dirac equation'' $D^* \phi =0$. In other words, the 
initially specified Dirac field $\chi$ (the ``Dirac field strength'') remains 
invariant under the following gauge transformations of its {\it Klein-Gordon potentials} 
(\ref{poten}): 
\be{gauge}  
\phi \mapsto \phi + \kappa, 
\ee
where $\kappa$ is an {\it arbitrary} solution to the Dirac equation
\be{dirac2}
D^*\kappa = 0 
\ee
Moreover, since for any $\kappa$ some Klein-Gordon potentials $\xi$ can be found, 
$\kappa = D\xi$, the gauge transformations (\ref{gauge}) may be represented in 
the familiar gradientlike form:
\be{grad}
\phi \mapsto \phi + D\xi
\ee
Now $\xi$ may be considered as an arbitrary field subject to the Klein-Gordon 
equation.  Thus we have proved that {\it the free Dirac field itself 
is a gauge invariant field resembling the Maxwell field strengths}. On the other 
hand, the Klein-Gordon field serves as the {\it field of potentials} with respect 
to the Dirac field. 
In this sense, these both cannot be regarded as independent fields 
describing different kinds of particles. As in the case of electromagnetic 
structures, these equations describe, in fact, the same physical field system in 
different representations and, up to a choice of gauge for the Klein-Gordon 
potentials, {\it the free Dirac and the 4-component Klein-Gordon equations are 
equivalent}, so that arbitrary  solution of the Dirac equation corresponds to some 
solution of the Klein-Gordon equation from the class of equivalence (\ref{grad}), 
and vice versa.

The gauge invariance of the Dirac equation, represented by the transformations 
(\ref{grad}), is defined by the gauge function $\xi(x)$ necessarily constrainted by 
the Klein-Gordon equation. Therefore, this type of gauge invariance differs from the 
ordinary gauge transformations of electromagnetic potentials
\be{emgauge}
A_\mu \mapsto A_\mu -\prt_\mu \alpha,
\ee
with $\alpha(x)$ an {\it arbitrary} smooth function of space-time coordinates 
$\{x_\mu\}$. However, if one requires, in addition, that the relativistic 
{\it Lorentz gauge equation} $\prt_\mu A^\mu=0$ be preserved under (\ref{emgauge}),  
then the gauge function $\alpha$ is known to satisfy {\it the d'Alembert wave 
equation} and is thus no longer arbitrary. The latter situation (in the  
{\it massless} case) is in a full analogy with that with the gauge freedom of the 
``massive'' Klein-Gordon potentials (\ref{grad}). It is also worth mentioning 
that a special variety of gauge structures, the so called ``weak'' gauge 
transformations with  parameters depending on coordinates only implicitly, i.e. 
through the initial field under transform, has been introduced in the framework of 
biquaternionic electrodynamics [16,17]. Of course, they are deeply 
related to the ``restricted'' gauge transformations (\ref{grad}) since if the 
gauge function $\xi$ depends only on the field $\xi=\xi(\phi(x))$ under transform, 
then it satisfies the Klein-Gordon equation, as required. Note that the ``weak'' gauge 
transformations constitute a proper subgroup of the full gauge group and are closely 
related to projective transformations in twistor space [17].

\section{Hierarchy of solutions to the Dirac and Klein-Gordon equations}

The above correspondence between the solutions of the Dirac and Klein-Gordon 
equations allows us to obtain (a chain of) derivative solutions to both equations. Let, say, 
a solution $\chi$ to the Dirac equation $D\chi=0$ be given. It also  
identically satisfies the Klein-Gordon equation 
\begin{equation}
0=D^*(D\chi) = (\Box-m^2)\chi \equiv 0.
\ee 
Consequently, the field $\chi_1$ defined as in (\ref{poten}) via derivatives of 
the initial field
\be{newsol}
\chi^\p = D^*\chi
\ee
again satisfies both the Dirac and Klein-Gordon equations:
\be{dirkg1}
D\chi^\p = D(D^*\chi) = (\Box-m^2)\chi \equiv 0,
\ee
\be{dirkg2}
(\Box-m^2)\chi^\p = D^*D\chi^\p \equiv 0.
\ee
However, since $D\chi=(i\gamma^\mu \prt_\mu -m)\chi = 0$, the new 
solution 
\be{propsol}
\chi^\p = D^*\chi = (i\gamma^\mu \prt_\mu +m)\chi = 2m\chi
\ee
is proportional to the old one. In order to obtain actually a functionally 
independent solution, one should make use of the {\it internal symmetry group} 
$\phi \mapsto P\phi,~P\in SL(4,\mathbb C)$ of the solutions $\phi$ to the 
Klein-Gordon equation (this symmetry relates also to the Lorentz invariance, 
see Section 4). Let us thus, instead of (\ref{newsol}), take 
\be{new1sol}
\chi_1 = D^*P_1\chi 
\ee    
with some arbitrary matrix $P_1\in SL(4,\mathbb C)$. 
Since $(\Box - m^2)P_1\chi=P_1(\Box - m^2)\chi=0$, the anew defined solution 
(\ref{new1sol}) will satisfy the Dirac equation, 
\be{newsat}
D\chi_1 = DD^*P_1\chi = 0. 
\ee
In the case when the matrix $P_1$ is not proportional to the $\gamma^5$ 
(in this case the solution $\chi_1$ is identically zero) or to the 
unit matrix, solution (\ref{new1sol}) will be functionally independent 
from the initial one. Continuing the procedure, one arrives at an (infinite) 
{\it chain of solutions} to the Dirac free equation of the following form: 
\be{chain}
\chi_N = D^*P_N...D^*P_2D^*P_1\chi, 
\ee
where, generally, all the matrices $P_i$ could be different. 
  
In order to imagine better the above procedure, one can first consider an  
analogous construction of hierarchy of solutions in the massless case, that 
is, to the {\it Weyl equation}.  In the spinor coordinates 
\be{spincoord}
u=t+z,~v=t-z,~w=x-iy,~p= x+iy 
\ee
($u,v$ being real and $w,p$ complex conjugated) the latter has the form of a 
set of two equations 
\be{weyl}
\prt_u \alpha = \prt_p \beta~~ (:=\gamma),~~\prt_w \alpha =\prt_v \beta~~ (:=\delta)
\ee
for the two components $\{\alpha,\beta\}$ of the Weyl 2-spinor. The d'Alembert 
equation manifests itself here as the compatibility condition for the 
Weyl system (\ref{weyl}):
\be{dalamb}
\Box \alpha = \frac{1}{4}(\prt_u\prt_v - \prt_w\prt_p) \alpha \equiv 0,
\ee
and analogously for the $\beta$-component. From (\ref{weyl}) it diectly follows 
that the derivative spinor $\{\delta,\gamma\}$ satisfies the ``reflected'' Weyl 
equation
\be{reflect}
\prt_u \delta = \prt_w \gamma~~(:=\pi),~~~\prt_p \delta =\prt_v \gamma~~(:=\rho),
\ee
(whereas the Weyl equation of initial type holds for the complex conjugated 2-spinor 
$\{\delta^*,\gamma^*\}$). On the other hand, for any solution to the Weyl equation 
there exist ``2-spinor potentials'' which  not only satisfy the d'Alembert 
equation but are themselves a new solution to the Weyl equation. Indeed, the system 
(\ref{weyl}) is solved locally by a new 2-spinor $\{\mu,\nu\}$ such that 
\be{weylpot}
\alpha = \prt_p  \mu,~\beta = \prt_u \mu; 
\ee
\be{weylpot2}
\alpha = \prt_v \nu,~\beta=\prt_w \nu,
\ee
so that the potential 2-spinor satisfies again the (reflected) Weyl equation 
\be{reflect2}
\prt_u \mu = \prt_w \nu,~~~\prt_p \mu =\prt_v \nu
\ee
Such procedure can be repeated and leads to a chain of 
solutions to the Weyl equation, to the d'Alembert equation and to the associated 
equations for complex Maxwell field (for details see [14]). On the other hand, 
if one resolves only, say, the first of the Weyl equations (\ref{weyl}) by means of the   
potential $\mu$ as in (\ref{weylpot}), then the second equation requires that the  
potential $\mu$ should satisfy the d'Alembert equation
\be{second}
\Box \mu = \frac{1}{4}(\prt_u\prt_v - \prt_w\prt_p) \mu = 0
\ee
Thus any Weyl 2-spinor can be obtained by derivation from a {\it d'Alembert   
potential}, i.e. from a solution to a one-component wave equation. 

Below we present some examples of the generating procedure in the 
massless and massive cases. Note that we do not deal here with rather 
a trivial case of regular wavelike solutions but, instead, consider solutions 
with a complicated structure of singularities. Let us first take the potential 
$\mu$ in (\ref{weylpot}) of the form 
\be{stereo}
\mu=\frac{p}{z+r}=\frac{x+iy}{z+r}=\tan\frac{\theta}{2}\exp(i\varphi),
\ee
which corresponds to the stereographic projection $S^2\mapsto \mathbb{C}$. It is easy 
to verify that this function satisfies the d'Alembert equation~\footnote{In addition, 
the potential (\ref{stereo}) satisfies the {\it complex eikonal equation} which is 
very important in the framework of {\it algebrodynamics}, see, e.g., [18,19]}. 
Then the 2-spinor $\{\alpha,\beta\}$ derived from (\ref{stereo}) in accordance with (\ref{weylpot}), 
\be{derived}
\alpha = \prt_p  \mu = \frac{1}{2r},~~\beta = \prt_u \mu= -\frac{\mu}{2r},  
\ee
identically satisfies the Weyl equation (\ref{weyl}), and for the derivative 
spinor $\{\delta,\gamma\}$ one gets from the latter:
\be{satis}
\delta = -\frac{p}{4r^3},~~\gamma=-\frac{z}{4r^3}.
\ee
The above spinor also satisfies the (reflected) Weyl 
equation (\ref{reflect}), and the corresponding derivatives determine the next 
2-spinor $\{\pi,\rho\}$ in the chain of solutions to the Weyl equation:
\be{next}
\pi=\frac{3zp}{8r^5},~~~\rho=\frac{r^2-3z^2}{8r^5}
\ee
It is not difficult to guess now that, in fact, this chain is directly related to 
the multipole expansion of the (complexified) electromagnetic potentials, the spinors 
$\{\alpha,\beta\},\{\delta,\gamma\}$ and $\{\pi,\rho\}$ representing  
the monopole, dipole and quadrupole terms, respectively. In more detail this 
correspondence will be presented elsewhere. 

Let us now return to the massive case and take, say, the following 
{\it stationary} solution to the Klein-Gordon free equation:
\be{fund}
\Phi = \frac{p}{r(z+r)}e^{-imt}=\frac{1}{r}\tan\frac{\theta}{2}
\exp{(i\varphi-\imath mt)}
\ee
where the ``frequency'' is necessarily equal to the Compton one: $\omega=m$. 
For the 4-component column of the Klein-Gordon potentials we trivially take
\be{col}
\phi^T=(\Phi,0,0,0).
\ee
Then, making use of the above-described procedure based on the expression 
(\ref{poten}) 
and allowing for generating the solutions to Dirac equation from the 
Klein-Gordon potentials, we easily obtain the following Dirac field $\chi$:
\be{bispinnext}
\chi^T=\left(\frac{2mp}{r(z+r)},0,-\frac{\imath p}{r^3},-\frac{\imath p^2(z+2r)}
{r^3(z+r)^2}\right)e^{-\imath mt}.
\ee

It is easy to check now that the soluion defined through (\ref{bispinnext}) as 
in (\ref{newsol}) is in fact proportional to (\ref{bispinnext}). If, however, 
one makes use of the recipe (\ref{new1sol}) and takes an auxiliary matrix 
$P_1=\gamma^3$ (this choice evidently preserves the Z-axial symmetry) then 
the following Ansatz results:
\be{ansatz}
\chi_1^T=(D^*\gamma^3\chi)^T = 
\left(\imath\frac{2mp}{r^3},0,\frac{6pz}{r^5},\frac{6 p^2}{r^5}
\right)e^{-\imath mt}.
\ee

One can see that the above field is indeed functionally independent from the 
initial one (\ref{bispinnext}) and obeys the Dirac free equation. Taking again, 
in accord with general prescription (\ref{chain}), 
\be{lastsol}
\chi_2=D^*\gamma^3\chi_1,~\chi_3=D^*\gamma^3\chi_2,~... ,
\ee
and so on, one can obtain other axisymmetrical solutions to the Dirac free 
equation and thus construct an hierarchy of these starting, in fact, 
from a single solution (\ref{fund}) to the Klein-Gordon free equation.

\section{Spinor transformations from scalar potentials}

Motivated by the above procedure, there naturally arises the problem of 
correspondence between the scalar nature of the Klein-Gordon potentials and the 
(bi)spinor type of transformations of the induced  Dirac field. Resolution 
of this problem is based on the existence of two {\it independent} symmetry 
groups of the 4-component Klein-Gordon equation, namely, of the Lorentz 
(Poincar\'e) space-time group and the internal symmetry group 
$\phi \mapsto P\phi,~~P\in SL(4,\mathbb{C})$ intermingling the components of 
the potentials. However, the situation is in fact much more interesting and 
complicated.

Indeed, in a ``4-rotated'' frame of reference, both the 4-component scalar 
field $\phi$ and the Klein-Gordon operator $(\Box-m^2)$ remain invariant. 
However, the first-order Dirac factors (\ref{factor}) are {\it not} invariant 
and transform as follows:
\be{dirac3}
(D^*)^\p=\gamma^\nu \prt^\p_\nu + m = \gamma^\nu L_\nu^\mu \prt_\mu + m,
\ee
where $\{L_\nu^\mu\}$ is the 6-parametric matrix of Lorentz tranformations.  
The new solution to the Dirac equation, generated, in accordance with (\ref{poten}), 
by the transformed operator
\be{trans}
\chi^\p = (D^*)^\p \phi , 
\ee
should be regarded as the {\it initial} Dirac field in the transformed frame of 
reference. On the other hand, the transformed Dirac operator (\ref{dirac3}) can be 
represented in the form 
\be{trdirac}
(D^*)^\p = SD^* S^{-1}, 
\ee
where the matrix $S\in SL(4,\mathbb{C})$ is now determined (up to a sign, $\pm S$) 
by the Lorentz transformations parameters that define the matrix $\{L_\nu^\mu\}$.
The above representation (\ref{trdirac}) is a 
direct consequence of a well known commutation property of the Dirac operator 
\be{commut} 
(D^*)^\p S = S D^*
\ee
responsible for the relativistic invariance of the Dirac equation in the  
generally accepted sense~\footnote{Indeed, if the Dirac field undergoes the ordinary 
(bi)spinor transformations, then one requires $(D^*)^\p \psi^\p = (D^*)^\p 
S\psi = S D^* \psi = 0$ to ensure the Dirac equation be form-invariant}. 

Thus the transformed solution to the Dirac equation in a 4-rotated frame takes 
the form 
\be{trdirac2}
\chi^\p = (SD^*S^{-1})\phi, ~~D^\p \chi^\p \equiv 0 .
\ee
However, these transformations, though linear in $\chi$, contain the derivatives of 
the fields, so that generally the new Dirac field cannot be represented in the form 
of a ``spintensorial'' transformation of the initial field $\chi$, 
\be{spinten}
\chi^\p \ne M\chi = M(D^*\phi)
\ee
with some ``representation matrix'' $M\in SL(4,\mathbb{C})$. Nonetheless, we 
have managed to generate transformed solutions to the Dirac equations in an 
arbitrary 4-rotated frame, which are in general different from the fields 
obtained by the canonical bispinor transformations.  Remarkably, in the above 
representation, we never meet any sort of two-valuedness of the transformed 
Dirac field. In particular, it is easly to see from (\ref{trdirac2}) that the 
field $\chi^\p$ evolves uniquely and continiously under the 3D rotations of 
coordinate frame, changes its sign under a rotation by $180^o$ and returns to 
its initial form (without change in sign!) under the complete turn by $360^o$.

It is now necessary to understand how one can restore the canonical bispinor 
transformations of the Dirac field. To do so, one should recall that, even in a 
fixed frame, the Dirac fields $\chi$ are defined non-uniquely since their 
Klein-Gordon potentials $\phi$ may be subject to transformations from the 
symmetry group:
\be{intsym}
\phi \mapsto P\phi,~~ \Rightarrow ~~\chi \mapsto D^* (P\phi)
\ee
with an {\it arbitrary} matrix $P\in SL(4,\mathbb{C})$. From a generic viewpoint, 
{\it all these Dirac fields should be regarded as (physically) equivalent!} 

Let us combine now the Dirac field transformations related to Lorentz 
rotations (with invariable potentials) (\ref{trdirac2}) with those related to 
alteration of potentials (\ref{intsym}) in accordance with their internal 
symmetry group. Then we arrive at the {\it equivalence class} of the Lorentz 
transformed Dirac fields
\be{eqclass}
\chi^\p_P =  (SD^*S^{-1})(P\phi). 
\ee
The last step one has to make, in order to obtain an explicit (though 
giving rise to the two-valuedness!) transformations law for the Dirac fields, 
is {\it to identify} two initially independent matrices: that of a space-time 
rotation $S$ and that of an internal intermingling of the potential components 
$P$, that is, to set $P\equiv S$.  
Then the transformed Dirac field (\ref{eqclass}) gets the familiar form
\be{canon}
\chi^\p = SD^*\phi \equiv S\chi. 
\ee
This is the canonical bispinor linear transformations law for the Dirac fields 
in terms of themselves. It does not contain any derivation and does not appeal 
to generating Klein-Gordon potentials. Nonetheless, one should not forget that 
this law is, in fact, not uniquely possible and even not the best one as 
compared, say, with the single-valued transformed Dirac field (\ref{trdirac2}).

\section{Conclusion} 

We have no opportunity here to discuss other peculiar problems related to the 
concept of Klein-Gordon potentials for the Dirac field. In particular, we 
postpone the discussion of a number of independent conservative energies, 
momenta, angular momenta and charges that may be ascribed to any solution of 
the associated Dirac and Klein-Gordon fields. The existence of such an ambiguity 
can, in principle, force one to reconsider the canonical quantization scheme 
for relativistic free fields based, in its considerable part, on the 
indefiniteness of energy density for the Dirac field and on its 
positive-definiteness for the Klein-Gordon field. 

We also put off a generalization of the above procedure to a Riemanninan 
space-time background or to the case when 
external, say, electromagnetic fields  are present. Since in these cases the 
ordinary Klein-Gordon operator can be no longer factorized into a product of 
Dirac-like operators, many features of the above connections between these fields 
are lost in the presence of external fields or on a curved background. However, 
the {\it squared Dirac equation} can then be used instead of the Klein-Gordon 
equation for potentials, and, from its solutions, one is able, as before, to 
derive a {\it whole chain} of solutions to the canonical 
Dirac equation with interactions. We are going to present the corresponding
examples in a forthcoming publication.

In any case, the obtained correspondence between the solutions and 
transformation properties of ``scalar''  and ``spinor'' fields looks rather 
unexpected and may shed new light on the classification of really independent 
fields and on possible ways of a unified description of bosons and fermions,  
alternative to those based on the supersymmetry hypothesis.  

In our construction, the Klein-Gordon and Dirac fields do not exist as two 
independent ones but manifest themselves as one and the same field in different 
representations. 
Specifically, the Klein-Gordon fields serve as potentials for the Dirac one. As 
for the question of what kind of particles actually corresponds to this field 
(if any!), an answer lies in the analysis of particular solutions and dynamical 
quantities that may be ascribed to them. In this respect, it seems intriguing 
that solutions to the free Dirac and Klein-Gordon equations are not at all 
exhausted by the regular plane wave solutions [8] but can possibly have a 
complicated structure of (point-like or extended) singularities. The simplest 
examples of such solution are represented by formulae 
(\ref{fund}) and (\ref{bispinnext}), (\ref{ansatz}) for the Klein-Gordon and 
Dirac equations, respectively.

\Acknow
{The author is grateful to the participants of the scientific seminar of the 
Russian Gravitational Society for a useful discussion and comments on the 
results presented in the article.}

\end{document}